\documentclass[3p,times,procedia,sort&compress]{elsarticle}
\usepackage{nupha_ecrc}
\usepackage{hyperref}

\volume{00}

\firstpage{1}

\journalname{Nuclear Physics A}

\runauth{M. Strickland}

\jid{nupha}

\jnltitlelogo{Nuclear Physics A}

\usepackage{amssymb}
\usepackage[figuresright]{rotating}

\begin{document}

\begin{frontmatter}



%% Instructions from Editor: Please use the following \dochead only in the preprint version (e-print arXiv etc.); 
%% use empty \dochead{} when submitting to Nuclear Physics A!
\dochead{XXVIIth International Conference on Ultrarelativistic Nucleus-Nucleus Collisions\\ (Quark Matter 2018)}
%% e.g. \dochead{17th International Conference on Dynamical Processes in Excited States of Solids}

\title{Small system studies:  A theory overview}


\author{Michael Strickland}

\address{Department of Physics, Kent State University, Kent, OH 44242 United States}

\begin{abstract}
There is now a substantial body of evidence that a deconfined quark-gluon plasma is created in ultrarelativistic collisions of heavy nuclei.  Some key observables which are used to gauge the production of the quark-gluon plasma are the hadronic spectra and multiplicities across many species, azimuthal anisotropies in the production of various hadrons, jet quenching, photon/dilepton production, and heavy quarkonium suppression.  A key question in the study of the quark-gluon plasma is what happens to these observables as one changes the system size.  In particular, it is very interesting to study collisions of ``small systems'' such as pp, dA, pA, and $e^+ e^-$ since naively one does not expect to produce a QGP in these cases.  One of the surprises from such studies was the existence of sizable azimuthal anisotropies in the hadron spectra in the highest multiplicity classes, which led to speculations that one can generate a QGP in such relatively rare events.  This interpretation, however, is complicated by the fact that there can be multiple sources of azimuthal anisotropy.  I will discuss our current understanding of the different sources of azimuthal anisotropies and the multiplicity windows in which each mechanism is expected to dominate.  I will also provide a critical discussion of the reliability of hydrodynamic models applied to small systems.  Finally, I will briefly discuss jet quenching and heavy quarkonium suppression in small systems and discuss whether or not these observables show any indication of the production of a small QGP droplet.
\end{abstract}


\end{frontmatter}

% \linenumbers

%%%%%%%%%%%%%%%%%%%%%%%%%%%%
\section*{}
%%%%%%%%%%%%%%%%%%%%%%%%%%%%

The main goal of the ongoing ultrarelativistic heavy ion collision program begin carried out at the Relativistic Heavy Ion Collider (RHIC) and the Large Hadron Collider (LHC) is to better understand the behavior of quantum chromodynamics (QCD) at high temperatures and densities.  Based on the asymptotic freedom of QCD, it was predicted decades ago that one could generate a deconfined ensemble of quarks and gluons by heating a hadronic system beyond temperatures on the order of the pion mass.  This expectation has been realized on the theoretical end by state-of-the-art lattice QCD and resummed perturbative QCD calculations which have determined that QCD, at small net baryon density, undergoes a smooth (crossover) phase transition from a system that is well-described as a hadron resonance gas to one that is well-described by quarks and gluons at a pseudo-critical temperature of approximately $T_c \sim 155$ MeV \cite{Borsanyi:2013bia,Bazavov:2014pvz,Haque:2014rua}.  On the experimental end, one finds that the low-momentum hadron spectra and collective flow observed in high-energy $AA$ collisions are well-described by dissipative hydrodynamic models that use the lattice QCD equation of state as input and other observables such as jet suppression and heavy quarkonium suppression are also well-described by models which presume the production of a hot and ``long lived'' quark-gluon plasma (QGP) with lifetimes $\tau \sim 10-15$ fm/c for Pb-Pb central collisions at the highest LHC beam energies.\footnote{For recent reviews, see Refs.~\cite{Jeon:2016uym,Romatschke:2017ejr,Florkowski:2017olj}}

In order to firmly establish that the collective flow observations are truly indicative of QGP production one must perform the same experiment but varying the system size and collision energy in order to see if similar features are observed in regions where one does not expect to generate a QGP.  Currently, there is an ongoing effort at RHIC to perform a beam energy scan in hopes of learning more about the behavior of hot matter at finite net baryon density in addition to studies of collisions of small systems such as pp, pA, and dA at RHIC and LHC energies, see e.g. \cite{Khachatryan:2010gv,Abelev:2012ola,CMS:2012qk,Chatrchyan:2013nka,Adare:2013piz,Aad:2014lta,Khachatryan:2015waa}.  For a recent longer review of small system studies, see Ref.~\cite{Nagle:2018nvi}.  One basic question these experiments are trying to ask is: Can we turn off the QGP?  In this proceedings contribution, I will present an overview of the work happening on the theoretical front.

\section{Azimuthal anisotropies in hadron production}
\label{sec:anisotropies}

%%%%%%%%%%%%%%%%%%%%%%%%%%%%%%%%%%%%%%%%%%%%%%%%%%%%%%%%%%%%%%%%
\begin{figure*}[t!]
\centerline{
\includegraphics[width=\linewidth]{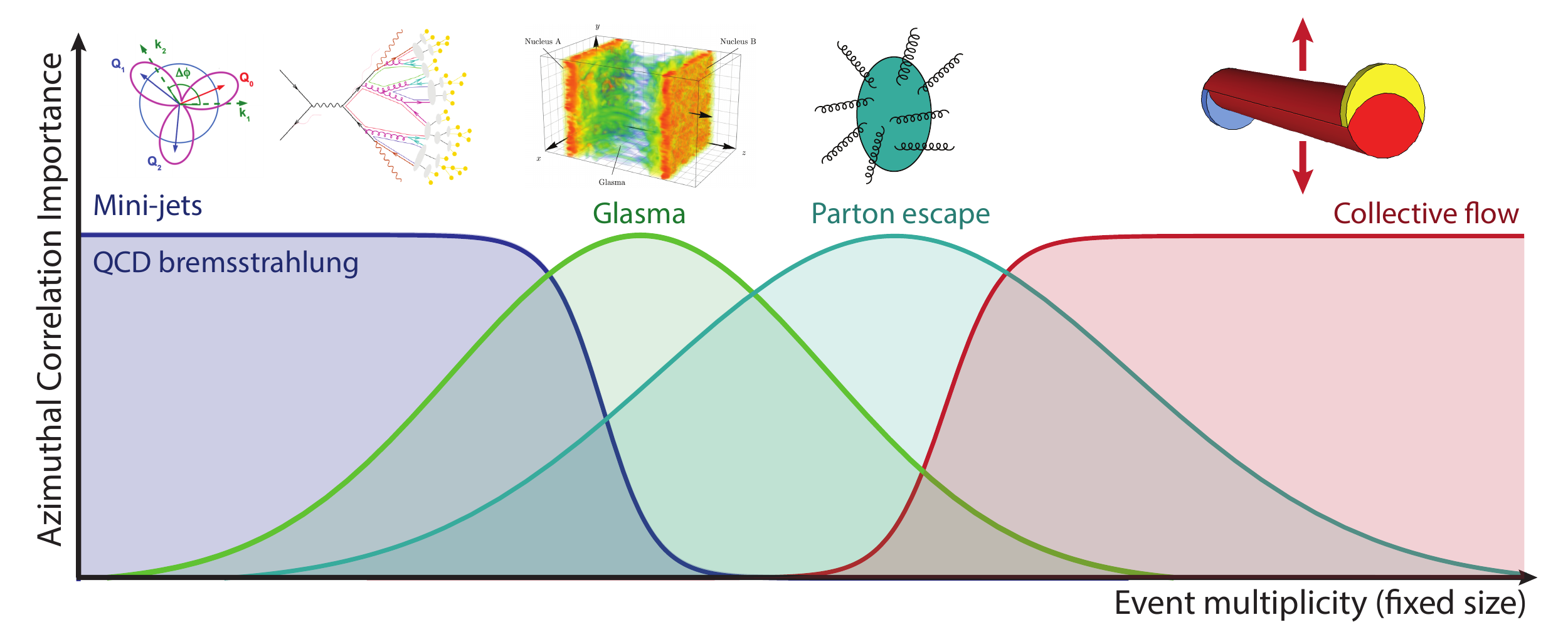}}
\caption{Cartoon depicting the various different sources of azimuthal anisotropy.  Height of each curve on the vertical axis is arbitrary.}
\label{fig1}
\end{figure*}
%%%%%%%%%%%%%%%%%%%%%%%%%%%%%%%%%%%%%%%%%%%%%%%%%%%%%%%%%%%%%%%%

Naively, one does not expect QGP production in small systems, however, experimentalists \cite{Khachatryan:2010gv,Abelev:2012ola,CMS:2012qk,Chatrchyan:2013nka,Adare:2013piz,Aad:2014lta,Khachatryan:2015waa} found that, in high-multiplicity events, there existed sizable azimuthal anisotropies in hadron production.  This was confusing because azimuthal anisotropies were thought to have their source in the transverse collective expansion of the system.  We have since learned that there can be multiple sources of the experimentally-observed azimuthal anisotropy.  In Fig.~\ref{fig1}, I illustrate the sources of azimuthal anisotropy which are expected to be important in each particle density window.  At the lowest multiplicities one expects parton bremsstrahlung, interference, and other effects  to dominate \cite{Kovner:2010xk,Ortiz:2013yxa,Gyulassy:2014cfa,Blok:2017pui}.   As one increases the multiplicity, it is expected that initial state correlations in the nuclear wavefunction \cite{Armesto:2006bv,Lappi:2009xa,Dumitru:2010iy,Dusling:2012iga,Dusling:2013qoz,Lappi:2015cgx,Schenke:2015aqa,Schenke:2016lrs,Dusling:2017dqg,Dusling:2017aot,Kovchegov:2018jun,Mace:2018yvl,Mace:2018vwq}, the parton escape mechanism \cite{Bzdak:2014dia,He:2015hfa,Li:2016ubw}, and collective flow driven by initial geometry aka ``hydrodynamic flow'' \cite{Bozek:2011if,Bozek:2013uha,Yan:2013laa,Romatschke:2015gxa,Shen:2016zpp,Weller:2017tsr,Mantysaari:2017cni} become the dominant mechanisms, respectively.  In all cases, theorists must be faced with experimental data which shows that, even in small systems, there is a hadronic mass ordering of the azimuthal anisotropies and an ordering of the flow coeficients with $v_2 > v_3 > v_4$ and approximately equal $v_{2n}$ for all $n \geq 2$.

\vspace{3mm}
\noindent
{\em Hydrodynamic models} --
The use of hydrodynamics to model small systems was pioneered by the Krakow group  \cite{Bozek:2011if,Bozek:2013uha} and since then multiple groups have successfully applied hydrodynamical models to the collision of small systems~\cite{Yan:2013laa,Romatschke:2015gxa,Shen:2016zpp,Weller:2017tsr,Mantysaari:2017cni}.  Despite the apparent success, the extension of hydrodynamic models to low multiplicities is fraught with challenges.  Firstly, traditional viscous hydrodynamic frameworks rely explicitly on an expansion in gradients of the fluid four-velocity.  As with all truncated expansions, if the magnitude of the expansion parameter (gradients) become large, the expansion itself becomes suspect.  This has motivated approaches that go to higher orders \cite{Jaiswal:2013vta} or perform a partial resummation of an infinite number of inverse Reynolds number \cite{Alqahtani:2017mhy}.\footnote{In cases where the gradient expansion has been extended to large order ($N \gtrsim 250$), one also finds that the series is an asymptotic series \cite{Florkowski:2017olj}.  While worrisome, in practice, an infinite number of gradients are resummed in the numerical evolution by promoting the shear tensor to be a dynamical field.  However, the manner in which gradients are resummed in the numerical evolution is sensitive to the order of the formal gradient expansion used to obtain the dissipative equations of motion.} In the context of second-order viscous hydrodynamics it is possible to estimate the maximum gradient which allows for a reliable hydrodynamic description.  This can be estimated by considering the analytic structure of retarded correlators \cite{Romatschke:2015gic,Grozdanov:2016vgg}.  In the low-momentum limit, one finds a weakly-damped mode which corresponds to the ``hydodynamic pole''.  Further down in the complex plane there can be additional poles or cuts which contribute to non-hydrodynamic behavior.  

If there is a large separation between the hydrodynamic pole and the non-hydrodynamic cuts/poles, then hydrodynamics is expected to be a reasonable approximation to the full dynamics, particularly at large times.  As the momentum is increased (smaller hotspots), the hydrodynamic pole becomes damped and moves down in the complex plane and no longer dominates the dynamics.  In the case of kinetic theory, one finds that at sufficiently high momentum, the hydrodynamic pole is `eaten' by a cut in the lower half plane \cite{Romatschke:2015gic}.  One finds with such an analysis that the smallest scale for a hotspot to be reliably described is on the order of 0.15 fm.  In the strong coupling limit, the non-hydrodynamic modes map to the quasinormal modes of the black hole in the bulk.  If the black hole is perturbed, there is a tower of modes excited which correspond to the `ring down' of the black hole.  In the strong coupling limit, the quasinormal modes ring down quickly, however, if one considers ${\cal N}=4$ SUSY Yang-Mills with large but finite t'Hooft coupling, one finds that the spacing between the quasinormal modes shrinks \cite{Grozdanov:2016vgg}.  Extrapolating this to intermediate coupling, one reaches a similar conclusion to the kinetic theory analysis regarding the scale at which the hydrodynamic approximation breaks down.  This is worrisome because models of fluctuating initial conditions can have small hotspots and associated shock waves (large gradients), casting doubt on the reliability of the equations of motion themselves.  One interesting recent proposal is to use kinetic theory evolution to describe the early-time large-gradient period of the evolution in order to bridge the gap between Glasma-like initial conditions and hydrodynamic evolution \cite{Kurkela:2018wud}.

Another issue faced by hydrodynamic modeling is that the dissipative equations of motion are typically derived by expanding around an isotropic thermal state.  If there are large non-equilibrium deviations, then once again the naive application of hydrodynamics becomes suspect.  This is particularly worrisome for the simulation of small systems because the system's lifetime is much shorter than that generated in AA collisions, even for the high-multiplicity events necessary.  Because of the short lifetime of the system, the system does not have sufficient time to become even approximately momentum-space isotropic in the local rest frame and one sees large gradients (Knudsen number) and inverse Reynolds number in a large hypervolume~\cite{Niemi:2014wta,Alqahtani:2016rth,Gallmeister:2018mcn}.  Additionally, because of the shorter lifetime the evolution will be more sensitive to non-hydrodynamic modes \cite{Strickland:2017kux}.  

In practice, one finds that hydrodynamic models applied to small systems, generate large non-equilibrium corrections that can, for example, drive the total pressure negative in a large hypervolume \cite{Mantysaari:2017cni}.  These large deviations additionally cause frequent violations of positivity of the one-particle distribution function on the freeze-out hypersurface \cite{Mantysaari:2017cni} which, without some scheme to regulate them, could result in a negative number of pions being produced, for example.  Groups deal with this problem differently.  For example, in SONIC \cite{Weller:2017tsr} an `exponentiation trick' introduced in Ref.~\cite{Pratt:2010jt} is used to guarantee positivity of the distribution.  This scheme reduces the effective size of the viscous corrections on the freeze-out hypersurface but is somewhat ad hoc. An alternative way to deal with the large viscous corrections is to use the anisotropic hydrodynamics (aHydro) framework which guarantees positivity of the one-particle distribution function at all points in spacetime, even far from equilibrium \cite{Alqahtani:2017mhy}.  This formalism has already been successfully applied to AA collisions at RHIC and LHC energies \cite{Alqahtani:2017jwl,Alqahtani:2017tnq,Almaalol:2018gjh}, so it is a natural next step to apply it to simulation of small systems.  

Despite these potential issues, phenomenological application of second-order viscous hydrodynamics has proceeded.  Surprisingly, one finds that, even given the large deviations from equilibrium and large corrections on the freeze-out hypersurface, hydrodynamic models can explain the observed $v_n$ down to $N_{\rm trk} \sim 70-100$.  A critical ingredient in such simulations is the use of partonic initial conditions in which nucleons are modelled as an ensemble of partons, see e.g. Refs.~\cite{Bozek:2016kpf,Welsh:2016siu,Mantysaari:2016how,Weller:2017tsr,Bozek:2017elk,Mantysaari:2017cni,Albacete:2017ajt}.  Without sub-nucleonic fluctuations, the initial geometric eccentricities in the transverse plane are not large enough to drive sufficient collective flow.  Viscous hydrodynamical models now include both shear and bulk viscosity, and subject to the caveats mentioned in the previous paragraphs, correctly reproduce the observed magnitude of $v_n$ in a variety of different collisions systems including the observed hadron mass ordering of the identified elliptic flow.  The quantitative success of hydrodynamical models strongly suggests that geometry-driven collective expansion is behind the observed azimuthal anisotropies at much lower multiplicities than one would naively expect.

\vspace{3mm}
\noindent
{\em Parton escape and kinetic theory approaches} --
Another mechanism which has been considered in the literature is the so-called parton escape mechanism \cite{Bzdak:2014dia,He:2015hfa,Li:2016ubw}.  In this case, azimuthal anisotropies are generated by differential escape from the deconfined partonic matter, with the escape probability depending on the angle of propagation through the system.  This mechanism and related `few hit' kinetic approaches \cite{Romatschke:2018wgi,Kurkela:2018qeb,Borghini:2018xum} are able to generate azimuthal anisotropies consistent with experimental observations.  Relatedly, partonic transport codes such as BAMPS \cite{Xu:2004mz,Gallmeister:2018mcn} and AMPT \cite{Lin:2004en} seem to also be able to reproduce the observed level of azimuthal anisotropy.  The fact that both hydrodynamic and kinetic theory models can describe the data may seem like a contradiction on the surface, however, I would like remind the reader that hydrodynamics can be thought of as a set of equations for the moments of the one-particle distribution function.  

Interpreted in this manner one can view the hydrodynamic evolution as an efficient approximation to the full kinetic evolution.  This suggests that we are seeing two sides of the same coin.  The success of quasiparticle-based hydrodynamical models which are derived explicitly from kinetic theory \cite{Alqahtani:2017jwl,Alqahtani:2017tnq,Almaalol:2018gjh} and quasiparticle-based QCD resummations of the equation of state at finite temperature and density \cite{Haque:2014rua} provide further evidence that one can think of the quark-gluon plasma as a massive quasiparticle gas.  This picture implicitly underlies finite temperature field theory applied to the QGP on many fronts, e.g. heavy quarkonium suppression, jet quenching, and the calculation of the.QGP shear viscosity.  The chief argument against this picture in the past has been the discrepancy between the observed $\eta/s$ and that predicted by leading-order perturbative QCD.  At this conference J. Ghiglieri presented a next-to-leading-order calculation of the shear viscosity to entropy density ratio of the QGP, with the results showing a significant reduction in $\eta/s$ bringing it into range  $\eta/s \sim 0.2$, albeit with a large uncertainty associated with both the scale variation and questions about the convergence of the series itself \cite{nloetas}.  If nothing else, taken together with the resummed equation of state results \cite{Haque:2014rua} this suggests that it may not be completely crazy to extrapolate our understanding of high-temperature QCD down to phenomenologically relevant temperatures.
 
\vspace{3mm}
\noindent
{\em Initial state correlation models} --
Another source of azimuthal anisotropy is initial state correlations present in the nuclear wavefunctions of the incoming nuclei.  These momentum correlations have their source in the inherent correlations associated with the QCD scale.  In the high energy limit chromoeletric and chromomagnetic fields are correlated in domains of size $L \sim Q_s^{-1}$ with $Q_s$ being the saturation scale.  Incoming partons see fields correlated on this length scale and are deflected in a coherent fashion by each domain.  Computing an ensemble average over the domain configurations, one finds that a non-vanishing long-range two-gluon correlation survives on an event-by-event basis~\cite{Armesto:2006bv,Dumitru:2010iy}. The resulting azimuthal anisotropies were originally understood analytically in terms of a subset of graphs called `Glasma-graphs'  \cite{Dumitru:2010iy,Dusling:2012iga,Dusling:2013qoz} which only produced even $v_n$ by symmetry.  Detailed numerical work showed that, if one went beyond the Glasma graphs, one could also produce odd $v_n$ \cite{Lappi:2009xa,Lappi:2015cgx,Schenke:2015aqa,Schenke:2016lrs}.  Subsequent work has show that it possible to generate non-vanishing $v_n\{m\}$ for all $n$ and $m$ using initial state correlations alone \cite{Dusling:2017dqg,Dusling:2017aot}.  This is particularly noteworthy since it had been claimed previously that one needed strong final state interactions in order to generate multiparticle correlations.  In addition, recently it was shown that it is possible to reproduce the observed mass ordering based on initial state correlations and subsequent gluon fragmentation using PYTHIA Lund string fragmentation \cite{Schenke:2016lrs}.  Once again, mass ordering which was thought to be a hallmark of hydrodynamic collective flow can be reproduced also with initial state sources of azimuthal anisotropy.  Finally, I mention in this context that recently Kovchegov and Skokov have demonstrated how to generate odd $v_n$ analytically by taking into account higher-order saturation corrections in the interactions of the nuclei and that the estimated magnitude of the odd harmonics is in the range of observations \cite{Kovchegov:2018jun}.  A numerical realization of this picture was presented at this conference by M. Mace~\cite{mace} and in a recent paper with collaborators \cite{Mace:2018vwq}.  The comparisons with data from the PHENIX collaboration show qualitative agreement with the trends seen in the experimental data for $v_2$ and $v_3$ versus $p_T$ in pA, dA, and ${}^3$He\,A collisions.

One obvious follow up question is to what extent do azimuthal anisotropies established during the initial stages of the collision survive subsequent final state interactions.  In this context, a recent paper by Greif et al presented at this conference provides the first study of the quantitative impact of each source for systems with different multiplicities \cite{Greif:2017bnr} .    To do this, the authors initialized the BAMPS transport code \cite{Xu:2004mz} with gluonic initial conditions coming from Glasma-induced correlations.  The authors found that at low multiplicity ($\langle dN_g/dy \rangle = 6$) initial state correlations survive with only small modification and that at high multiplicity ($\langle dN_g/dy \rangle = 26$) final state interactions dominate.  In the high multiplicity class, they authors found that initial state correlations in the gluon distribution resulted in an approximately 25\% modification of model predictions for $v_2\{2\}$ with the importance increasing as the multiplicity was decreased.  This study allows us to start to attach numbers to the cartoon shown in Fig.~\ref{fig1}, further bolstering the case that, for $N_{\rm trk} \gtrsim 100$, the azimuthal anisotropies observed are dominated by final state interactions.

As a final note concerning initial state correlations, I mention that some very challenging results were presented by the ALEPH and ZEUS and collaborations at this conference \cite{aleph,zeus}.  The ALEPH collaboration reported that they found no long-range correlations in rapidity in $e^+ e^-$ collisions and that their results were well-explained by Pythia 6.1 without final state interactions.  The ZEUS collaboration reported that they found that the correlation coefficients $c_{2,3,4}\{2\}$ were all consistent with zero in $e^- p$ collisions and, once again, that the results were consistent with predictions from existing Monte-Carlo generators.  The $e^- p$ results are particularly interesting because one would naively expect that the standard Glasma-based dilute-dense picture would apply  in this case.  It is possible that the large rapidity separation cut ($\Delta \eta > 2$) pushes one into the BFKL ladder regime which decorrelates the gluons, however, if this is true, then the same would apply to $p p$ and $p A$ collisions.  At this point, the ZEUS finding presents an existential problem for initial state correlations which should be further studied by both theorists and experimentalists.

\vspace{3mm}
\noindent
{\em Parton bremsstrahlung, interference, and other effects} --
Finally, I mention that at very low multiplicities one expects vacuum QCD effects to dominate, which include partonic bremsstrahlung, interference, color reconnection, dipole orientation, etc:  \cite{Kovner:2010xk,Ortiz:2013yxa,Gyulassy:2014cfa,Blok:2017pui}.  Works along this direction are important and ongoing, however, it seems clear that at higher multiplicities such effects would be wiped out by final state interactions.

\section{Jet quenching and heavy quarkonium suppression}

In the remaining space, I would like to briefly mention that the story is bigger than azimuthal anisotropies.  They have received a lot of attention because of the confusion caused by the fact they existed also for small systems, however, firm evidence of generation of a QGP relies on the existence of multiple signatures.  For example,  in AA collisions we have observed sizable jet quenching associated with partonic energy loss in the QGP, the level of which cannot be explained by purely hadronic mechanisms.  Is such a signature present in high-multiplicity collisions of small systems?  On the theoretical front, one expects there to be some effect, however, due to the small size and short lifetime of the QGP generated in collisions of small systems, the effect is estimated to be significantly smaller than what is observed in AA collisions.  In Ref.~\cite{Shen:2016egw}, for example, the authors estimated that, in the 0-1\% centrality class, $R_{pA}$ would show a maximal 25\% deviation from unity at $p_T = 10$ GeV, with the effect decreasing at high $p_T$.  Experimentally it is very difficult, if not impossible, to get sufficient unbiased statistics in such extreme centrality classes.  A recent study by the ALICE collaboration which measured jet quenching in pA collisions in the 0-10\% centrality class found that at $p_{T, \rm jet} = 57$ GeV one has $R_{CP} = 1.09 \pm 0.02{\rm (stat)} \pm 0.06{\rm (sys)}$, which indicates enhancement instead of suppression but, within error bars, is consistent with no suppression \cite{Acharya:2017okq}.  It will be interesting to see if future experimental analyses can be extended to even more central centrality classes, so that we might more quantitatively assess whether not jet suppression is consistent with the production of a tiny droplet of QGP in small systems.

Finally, I would like to discuss the possibility to perform a similar analysis using the suppression of bottomonium.  In high-energy AA collisions, bottomonium suppression is a particularly nice `smoking gun' for QGP creation since cold nuclear matter effects are expected to be small in Upsilon production.  Experimental observations of bottomonium suppression in pA collisions indicate that at central rapidity and average $p_T$, $R_{AA} \sim 0.7 - 0.8$, while in AA collisions one finds $R_{AA} \sim 0.35$ in central collisions \cite{Khachatryan:2016xxp,CMS5TeV,ALICE5TeV,Aaboud:2017cif}.  The pA observations are consistent with cold nuclear matter effects while the AA observations can thus far only be explained by models that have strong final state interactions, see e.g. \cite{Emerick:2011xu,Strickland:2011mw,Strickland:2011aa,Krouppa:2015yoa,Krouppa:2016jcl,Du:2017qkv,Krouppa:2017lsw,Krouppa:2017jlg,Ferreiro:2018wbd}.  Again, it is challenging, if not impossible, to collect enough statistics on bottomonium suppression in high-multiplicity events, however, we should definitely include this on a wish list for the future.

\section{Conclusions and outlook}

In conclusion, we return to the basic question asked in the introduction:  Can we turn off the QGP?  While this may have been confusing a few years ago, I would venture to say that the answer is a definitive yes.  The confusion mostly stemmed from the community's interpretation of observed azimuthal anisotropies as evidence for true hydrodynamic flow.  We now understand that there are multiple sources of azimuthal anisotropy which play important roles in different multiplicity windows and on the theoretical front there is progress in estimating the relative importance of each of these mechanisms.  Based on the findings to date, I would estimate that the azimuthal anisotropies observed are consistent with strong final state interactions and true collective flow only for $N_{\rm trk} \gtrsim 100$.  Despite the lower bound, this is a quite surprising result which indicates that dissipative hydrodynamical models might be applicable at much lower multiplicities/energies than anyone naively expected.  Of course, as mentioned herein, these hydrodynamical studies are not without issues and future work is needed to extend existing dissipative hydrodynamic frameworks to situations which are quite far from equilibrium.  Finally, on the hard probes front, based on experimental data released to date, there is no indication of QGP generation in collisions of small systems, however, these studies are extremely statistics limited making it difficult to study the signatures.  That said, any further experimental information on this front would be very welcome.

\vspace{4mm}

\noindent
{\bf Acknowledgments} -- M. Strickland was supported by the U.S. Department of Energy, Office of Science, Office of Nuclear Physics under Award No. DE-SC0013470.

\vspace{-2mm}

\bibliographystyle{model1a-num-names}
\bibliography{strickland}

\begin{thebibliography}{84}
\expandafter\ifx\csname natexlab\endcsname\relax\def\natexlab#1{#1}\fi
\providecommand{\url}[1]{\texttt{#1}}
\providecommand{\href}[2]{#2}
\providecommand{\path}[1]{#1}
\providecommand{\DOIprefix}{doi:}
\providecommand{\ArXivprefix}{arXiv:}
\providecommand{\URLprefix}{URL: }
\providecommand{\Pubmedprefix}{pmid:}
\providecommand{\doi}[1]{\href{http://dx.doi.org/#1}{\path{#1}}}
\providecommand{\Pubmed}[1]{\href{pmid:#1}{\path{#1}}}
\providecommand{\bibinfo}[2]{#2}
\ifx\xfnm\relax \def\xfnm[#1]{\unskip,\space#1}\fi
%Type = Article
\bibitem[{Borsanyi et~al.(2014)Borsanyi, Fodor, Hoelbling, Katz, Krieg, and
  Szabo}]{Borsanyi:2013bia}
\bibinfo{author}{S.~Borsanyi}, \bibinfo{author}{Z.~Fodor},
  \bibinfo{author}{C.~Hoelbling}, \bibinfo{author}{S.~D. Katz},
  \bibinfo{author}{S.~Krieg}, \bibinfo{author}{K.~K. Szabo},
  \bibinfo{journal}{Phys. Lett.} \bibinfo{volume}{B730} (\bibinfo{year}{2014})
  \bibinfo{pages}{99--104}. \href{http://arxiv.org/abs/1309.5258}{\tt
  arXiv:1309.5258}.
%Type = Article
\bibitem[{Bazavov et~al.(2014)}]{Bazavov:2014pvz}
\bibinfo{author}{A.~Bazavov}, et~al. (\bibinfo{collaboration}{HotQCD}),
  \bibinfo{journal}{Phys. Rev.} \bibinfo{volume}{D90} (\bibinfo{year}{2014})
  \bibinfo{pages}{094503}. \href{http://arxiv.org/abs/1407.6387}{\tt
  arXiv:1407.6387}.
%Type = Article
\bibitem[{Haque et~al.(2014)Haque, Bandyopadhyay, Andersen, Mustafa,
  Strickland, and Su}]{Haque:2014rua}
\bibinfo{author}{N.~Haque}, \bibinfo{author}{A.~Bandyopadhyay},
  \bibinfo{author}{J.~O. Andersen}, \bibinfo{author}{M.~G. Mustafa},
  \bibinfo{author}{M.~Strickland}, \bibinfo{author}{N.~Su},
  \bibinfo{journal}{JHEP} \bibinfo{volume}{05} (\bibinfo{year}{2014})
  \bibinfo{pages}{027}. \href{http://arxiv.org/abs/1402.6907}{\tt
  arXiv:1402.6907}.
%Type = Incollection
\bibitem[{Jeon and Heinz(2016)}]{Jeon:2016uym}
\bibinfo{author}{S.~Jeon}, \bibinfo{author}{U.~Heinz}, in:
  \bibinfo{editor}{X.-N. Wang} (Ed.), \bibinfo{booktitle}{Quark-Gluon Plasma
  5}, \bibinfo{year}{2016}, pp. \bibinfo{pages}{131--187}.
%Type = Article
\bibitem[{Romatschke and Romatschke(2017)}]{Romatschke:2017ejr}
\bibinfo{author}{P.~Romatschke}, \bibinfo{author}{U.~Romatschke}
  (\bibinfo{year}{2017}). \href{http://arxiv.org/abs/1712.05815}{\tt
  arXiv:1712.05815}.
%Type = Article
\bibitem[{Florkowski et~al.(2018)Florkowski, Heller, and
  Spalinski}]{Florkowski:2017olj}
\bibinfo{author}{W.~Florkowski}, \bibinfo{author}{M.~P. Heller},
  \bibinfo{author}{M.~Spalinski}, \bibinfo{journal}{Rept. Prog. Phys.}
  \bibinfo{volume}{81} (\bibinfo{year}{2018}) \bibinfo{pages}{046001}.
  \href{http://arxiv.org/abs/1707.02282}{\tt arXiv:1707.02282}.
%Type = Article
\bibitem[{Khachatryan et~al.(2010)}]{Khachatryan:2010gv}
\bibinfo{author}{V.~Khachatryan}, et~al. (\bibinfo{collaboration}{CMS}),
  \bibinfo{journal}{JHEP} \bibinfo{volume}{09} (\bibinfo{year}{2010})
  \bibinfo{pages}{091}. \href{http://arxiv.org/abs/1009.4122}{\tt
  arXiv:1009.4122}.
%Type = Article
\bibitem[{Abelev et~al.(2013)}]{Abelev:2012ola}
\bibinfo{author}{B.~Abelev}, et~al. (\bibinfo{collaboration}{ALICE}),
  \bibinfo{journal}{Phys. Lett.} \bibinfo{volume}{B719} (\bibinfo{year}{2013})
  \bibinfo{pages}{29--41}. \href{http://arxiv.org/abs/1212.2001}{\tt
  arXiv:1212.2001}.
%Type = Article
\bibitem[{Chatrchyan et~al.(2013{\natexlab{a}})}]{CMS:2012qk}
\bibinfo{author}{S.~Chatrchyan}, et~al. (\bibinfo{collaboration}{CMS}),
  \bibinfo{journal}{Phys. Lett.} \bibinfo{volume}{B718}
  (\bibinfo{year}{2013}{\natexlab{a}}) \bibinfo{pages}{795--814}.
  \href{http://arxiv.org/abs/1210.5482}{\tt arXiv:1210.5482}.
%Type = Article
\bibitem[{Chatrchyan et~al.(2013{\natexlab{b}})}]{Chatrchyan:2013nka}
\bibinfo{author}{S.~Chatrchyan}, et~al. (\bibinfo{collaboration}{CMS}),
  \bibinfo{journal}{Phys. Lett.} \bibinfo{volume}{B724}
  (\bibinfo{year}{2013}{\natexlab{b}}) \bibinfo{pages}{213--240}.
  \href{http://arxiv.org/abs/1305.0609}{\tt arXiv:1305.0609}.
%Type = Article
\bibitem[{Adare et~al.(2013)}]{Adare:2013piz}
\bibinfo{author}{A.~Adare}, et~al. (\bibinfo{collaboration}{PHENIX}),
  \bibinfo{journal}{Phys. Rev. Lett.} \bibinfo{volume}{111}
  (\bibinfo{year}{2013}) \bibinfo{pages}{212301}.
  \href{http://arxiv.org/abs/1303.1794}{\tt arXiv:1303.1794}.
%Type = Article
\bibitem[{Aad et~al.(2014)}]{Aad:2014lta}
\bibinfo{author}{G.~Aad}, et~al. (\bibinfo{collaboration}{ATLAS}),
  \bibinfo{journal}{Phys. Rev.} \bibinfo{volume}{C90} (\bibinfo{year}{2014})
  \bibinfo{pages}{044906}. \href{http://arxiv.org/abs/1409.1792}{\tt
  arXiv:1409.1792}.
%Type = Article
\bibitem[{Khachatryan et~al.(2015)}]{Khachatryan:2015waa}
\bibinfo{author}{V.~Khachatryan}, et~al. (\bibinfo{collaboration}{CMS}),
  \bibinfo{journal}{Phys. Rev. Lett.} \bibinfo{volume}{115}
  (\bibinfo{year}{2015}) \bibinfo{pages}{012301}.
  \href{http://arxiv.org/abs/1502.05382}{\tt arXiv:1502.05382}.
%Type = Article
\bibitem[{Nagle and Zajc(2018)}]{Nagle:2018nvi}
\bibinfo{author}{J.~L. Nagle}, \bibinfo{author}{W.~A. Zajc}
  (\bibinfo{year}{2018}). \href{http://arxiv.org/abs/1801.03477}{\tt
  arXiv:1801.03477}.
%Type = Article
\bibitem[{Kovner and Lublinsky(2011)}]{Kovner:2010xk}
\bibinfo{author}{A.~Kovner}, \bibinfo{author}{M.~Lublinsky},
  \bibinfo{journal}{Phys. Rev.} \bibinfo{volume}{D83} (\bibinfo{year}{2011})
  \bibinfo{pages}{034017}. \href{http://arxiv.org/abs/1012.3398}{\tt
  arXiv:1012.3398}.
%Type = Article
\bibitem[{Ortiz~Velasquez et~al.(2013)Ortiz~Velasquez, Christiansen,
  Cuautle~Flores, Maldonado~Cervantes, and Pai?}]{Ortiz:2013yxa}
\bibinfo{author}{A.~Ortiz~Velasquez}, \bibinfo{author}{P.~Christiansen},
  \bibinfo{author}{E.~Cuautle~Flores},
  \bibinfo{author}{I.~Maldonado~Cervantes}, \bibinfo{author}{G.~Pai?},
  \bibinfo{journal}{Phys. Rev. Lett.} \bibinfo{volume}{111}
  (\bibinfo{year}{2013}) \bibinfo{pages}{042001}.
  \href{http://arxiv.org/abs/1303.6326}{\tt arXiv:1303.6326}.
%Type = Article
\bibitem[{Gyulassy et~al.(2014)Gyulassy, Levai, Vitev, and
  Biro}]{Gyulassy:2014cfa}
\bibinfo{author}{M.~Gyulassy}, \bibinfo{author}{P.~Levai},
  \bibinfo{author}{I.~Vitev}, \bibinfo{author}{T.~S. Biro},
  \bibinfo{journal}{Phys. Rev.} \bibinfo{volume}{D90} (\bibinfo{year}{2014})
  \bibinfo{pages}{054025}. \href{http://arxiv.org/abs/1405.7825}{\tt
  arXiv:1405.7825}.
%Type = Article
\bibitem[{Blok et~al.(2017)Blok, Jäkel, Strikman, and
  Wiedemann}]{Blok:2017pui}
\bibinfo{author}{B.~Blok}, \bibinfo{author}{C.~D. Jäkel},
  \bibinfo{author}{M.~Strikman}, \bibinfo{author}{U.~A. Wiedemann},
  \bibinfo{journal}{JHEP} \bibinfo{volume}{12} (\bibinfo{year}{2017})
  \bibinfo{pages}{074}. \href{http://arxiv.org/abs/1708.08241}{\tt
  arXiv:1708.08241}.
%Type = Article
\bibitem[{Armesto et~al.(2007)Armesto, McLerran, and Pajares}]{Armesto:2006bv}
\bibinfo{author}{N.~Armesto}, \bibinfo{author}{L.~McLerran},
  \bibinfo{author}{C.~Pajares}, \bibinfo{journal}{Nucl. Phys.}
  \bibinfo{volume}{A781} (\bibinfo{year}{2007}) \bibinfo{pages}{201--208}.
  \href{http://arxiv.org/abs/hep-ph/0607345}{\tt arXiv:hep-ph/0607345}.
%Type = Article
\bibitem[{Lappi et~al.(2010)Lappi, Srednyak, and Venugopalan}]{Lappi:2009xa}
\bibinfo{author}{T.~Lappi}, \bibinfo{author}{S.~Srednyak},
  \bibinfo{author}{R.~Venugopalan}, \bibinfo{journal}{JHEP}
  \bibinfo{volume}{01} (\bibinfo{year}{2010}) \bibinfo{pages}{066}.
  \href{http://arxiv.org/abs/0911.2068}{\tt arXiv:0911.2068}.
%Type = Article
\bibitem[{Dumitru et~al.(2011)Dumitru, Dusling, Gelis, Jalilian-Marian, Lappi,
  and Venugopalan}]{Dumitru:2010iy}
\bibinfo{author}{A.~Dumitru}, \bibinfo{author}{K.~Dusling},
  \bibinfo{author}{F.~Gelis}, \bibinfo{author}{J.~Jalilian-Marian},
  \bibinfo{author}{T.~Lappi}, \bibinfo{author}{R.~Venugopalan},
  \bibinfo{journal}{Phys. Lett.} \bibinfo{volume}{B697} (\bibinfo{year}{2011})
  \bibinfo{pages}{21--25}. \href{http://arxiv.org/abs/1009.5295}{\tt
  arXiv:1009.5295}.
%Type = Article
\bibitem[{Dusling and Venugopalan(2012)}]{Dusling:2012iga}
\bibinfo{author}{K.~Dusling}, \bibinfo{author}{R.~Venugopalan},
  \bibinfo{journal}{Phys. Rev. Lett.} \bibinfo{volume}{108}
  (\bibinfo{year}{2012}) \bibinfo{pages}{262001}.
  \href{http://arxiv.org/abs/1201.2658}{\tt arXiv:1201.2658}.
%Type = Article
\bibitem[{Dusling and Venugopalan(2013)}]{Dusling:2013qoz}
\bibinfo{author}{K.~Dusling}, \bibinfo{author}{R.~Venugopalan},
  \bibinfo{journal}{Phys. Rev.} \bibinfo{volume}{D87} (\bibinfo{year}{2013})
  \bibinfo{pages}{094034}. \href{http://arxiv.org/abs/1302.7018}{\tt
  arXiv:1302.7018}.
%Type = Article
\bibitem[{Lappi(2016)}]{Lappi:2015cgx}
\bibinfo{author}{T.~Lappi}, \bibinfo{journal}{Nucl. Phys.}
  \bibinfo{volume}{A956} (\bibinfo{year}{2016}) \bibinfo{pages}{537--540}.
  \href{http://arxiv.org/abs/1512.07209}{\tt arXiv:1512.07209}.
%Type = Article
\bibitem[{Schenke et~al.(2015)Schenke, Schlichting, and
  Venugopalan}]{Schenke:2015aqa}
\bibinfo{author}{B.~Schenke}, \bibinfo{author}{S.~Schlichting},
  \bibinfo{author}{R.~Venugopalan}, \bibinfo{journal}{Phys. Lett.}
  \bibinfo{volume}{B747} (\bibinfo{year}{2015}) \bibinfo{pages}{76--82}.
  \href{http://arxiv.org/abs/1502.01331}{\tt arXiv:1502.01331}.
%Type = Article
\bibitem[{Schenke et~al.(2016)Schenke, Schlichting, Tribedy, and
  Venugopalan}]{Schenke:2016lrs}
\bibinfo{author}{B.~Schenke}, \bibinfo{author}{S.~Schlichting},
  \bibinfo{author}{P.~Tribedy}, \bibinfo{author}{R.~Venugopalan},
  \bibinfo{journal}{Phys. Rev. Lett.} \bibinfo{volume}{117}
  (\bibinfo{year}{2016}) \bibinfo{pages}{162301}.
  \href{http://arxiv.org/abs/1607.02496}{\tt arXiv:1607.02496}.
%Type = Article
\bibitem[{Dusling et~al.(2018{\natexlab{a}})Dusling, Mace, and
  Venugopalan}]{Dusling:2017dqg}
\bibinfo{author}{K.~Dusling}, \bibinfo{author}{M.~Mace},
  \bibinfo{author}{R.~Venugopalan}, \bibinfo{journal}{Phys. Rev. Lett.}
  \bibinfo{volume}{120} (\bibinfo{year}{2018}{\natexlab{a}})
  \bibinfo{pages}{042002}. \href{http://arxiv.org/abs/1705.00745}{\tt
  arXiv:1705.00745}.
%Type = Article
\bibitem[{Dusling et~al.(2018{\natexlab{b}})Dusling, Mace, and
  Venugopalan}]{Dusling:2017aot}
\bibinfo{author}{K.~Dusling}, \bibinfo{author}{M.~Mace},
  \bibinfo{author}{R.~Venugopalan}, \bibinfo{journal}{Phys. Rev.}
  \bibinfo{volume}{D97} (\bibinfo{year}{2018}{\natexlab{b}})
  \bibinfo{pages}{016014}. \href{http://arxiv.org/abs/1706.06260}{\tt
  arXiv:1706.06260}.
%Type = Article
\bibitem[{Kovchegov and Skokov(2018)}]{Kovchegov:2018jun}
\bibinfo{author}{Y.~V. Kovchegov}, \bibinfo{author}{V.~V. Skokov},
  \bibinfo{journal}{Phys. Rev.} \bibinfo{volume}{D97} (\bibinfo{year}{2018})
  \bibinfo{pages}{094021}. \href{http://arxiv.org/abs/1802.08166}{\tt
  arXiv:1802.08166}.
%Type = Article
\bibitem[{Mace et~al.(2018{\natexlab{a}})Mace, Skokov, Tribedy, and
  Venugopalan}]{Mace:2018yvl}
\bibinfo{author}{M.~Mace}, \bibinfo{author}{V.~V. Skokov},
  \bibinfo{author}{P.~Tribedy}, \bibinfo{author}{R.~Venugopalan}
  (\bibinfo{year}{2018}{\natexlab{a}}).
  \href{http://arxiv.org/abs/1807.00825}{\tt arXiv:1807.00825}.
%Type = Article
\bibitem[{Mace et~al.(2018{\natexlab{b}})Mace, Skokov, Tribedy, and
  Venugopalan}]{Mace:2018vwq}
\bibinfo{author}{M.~Mace}, \bibinfo{author}{V.~V. Skokov},
  \bibinfo{author}{P.~Tribedy}, \bibinfo{author}{R.~Venugopalan},
  \bibinfo{journal}{Phys. Rev. Lett.} \bibinfo{volume}{121}
  (\bibinfo{year}{2018}{\natexlab{b}}) \bibinfo{pages}{052301}.
  \href{http://arxiv.org/abs/1805.09342}{\tt arXiv:1805.09342}.
%Type = Article
\bibitem[{Bzdak and Ma(2014)}]{Bzdak:2014dia}
\bibinfo{author}{A.~Bzdak}, \bibinfo{author}{G.-L. Ma}, \bibinfo{journal}{Phys.
  Rev. Lett.} \bibinfo{volume}{113} (\bibinfo{year}{2014})
  \bibinfo{pages}{252301}. \href{http://arxiv.org/abs/1406.2804}{\tt
  arXiv:1406.2804}.
%Type = Article
\bibitem[{He et~al.(2016)He, Edmonds, Lin, Liu, Molnar, and Wang}]{He:2015hfa}
\bibinfo{author}{L.~He}, \bibinfo{author}{T.~Edmonds}, \bibinfo{author}{Z.-W.
  Lin}, \bibinfo{author}{F.~Liu}, \bibinfo{author}{D.~Molnar},
  \bibinfo{author}{F.~Wang}, \bibinfo{journal}{Phys. Lett.}
  \bibinfo{volume}{B753} (\bibinfo{year}{2016}) \bibinfo{pages}{506--510}.
  \href{http://arxiv.org/abs/1502.05572}{\tt arXiv:1502.05572}.
%Type = Article
\bibitem[{Li et~al.(2017)Li, He, Lin, Molnar, Wang, and Xie}]{Li:2016ubw}
\bibinfo{author}{H.~Li}, \bibinfo{author}{L.~He}, \bibinfo{author}{Z.-W. Lin},
  \bibinfo{author}{D.~Molnar}, \bibinfo{author}{F.~Wang},
  \bibinfo{author}{W.~Xie}, \bibinfo{journal}{Phys. Rev.} \bibinfo{volume}{C96}
  (\bibinfo{year}{2017}) \bibinfo{pages}{014901}.
  \href{http://arxiv.org/abs/1604.07387}{\tt arXiv:1604.07387}.
%Type = Article
\bibitem[{Bozek(2012)}]{Bozek:2011if}
\bibinfo{author}{P.~Bozek}, \bibinfo{journal}{Phys. Rev.} \bibinfo{volume}{C85}
  (\bibinfo{year}{2012}) \bibinfo{pages}{014911}.
  \href{http://arxiv.org/abs/1112.0915}{\tt arXiv:1112.0915}.
%Type = Article
\bibitem[{Bozek and Broniowski(2013)}]{Bozek:2013uha}
\bibinfo{author}{P.~Bozek}, \bibinfo{author}{W.~Broniowski},
  \bibinfo{journal}{Phys. Rev.} \bibinfo{volume}{C88} (\bibinfo{year}{2013})
  \bibinfo{pages}{014903}. \href{http://arxiv.org/abs/1304.3044}{\tt
  arXiv:1304.3044}.
%Type = Article
\bibitem[{Yan and Ollitrault(2014)}]{Yan:2013laa}
\bibinfo{author}{L.~Yan}, \bibinfo{author}{J.-Y. Ollitrault},
  \bibinfo{journal}{Phys. Rev. Lett.} \bibinfo{volume}{112}
  (\bibinfo{year}{2014}) \bibinfo{pages}{082301}.
  \href{http://arxiv.org/abs/1312.6555}{\tt arXiv:1312.6555}.
%Type = Article
\bibitem[{Romatschke(2015)}]{Romatschke:2015gxa}
\bibinfo{author}{P.~Romatschke}, \bibinfo{journal}{Eur. Phys. J.}
  \bibinfo{volume}{C75} (\bibinfo{year}{2015}) \bibinfo{pages}{305}.
  \href{http://arxiv.org/abs/1502.04745}{\tt arXiv:1502.04745}.
%Type = Article
\bibitem[{Shen et~al.(2017)Shen, Paquet, Denicol, Jeon, and
  Gale}]{Shen:2016zpp}
\bibinfo{author}{C.~Shen}, \bibinfo{author}{J.-F. Paquet},
  \bibinfo{author}{G.~S. Denicol}, \bibinfo{author}{S.~Jeon},
  \bibinfo{author}{C.~Gale}, \bibinfo{journal}{Phys. Rev.}
  \bibinfo{volume}{C95} (\bibinfo{year}{2017}) \bibinfo{pages}{014906}.
  \href{http://arxiv.org/abs/1609.02590}{\tt arXiv:1609.02590}.
%Type = Article
\bibitem[{Weller and Romatschke(2017)}]{Weller:2017tsr}
\bibinfo{author}{R.~D. Weller}, \bibinfo{author}{P.~Romatschke},
  \bibinfo{journal}{Phys. Lett.} \bibinfo{volume}{B774} (\bibinfo{year}{2017})
  \bibinfo{pages}{351--356}. \href{http://arxiv.org/abs/1701.07145}{\tt
  arXiv:1701.07145}.
%Type = Article
\bibitem[{Mäntysaari et~al.(2017)Mäntysaari, Schenke, Shen, and
  Tribedy}]{Mantysaari:2017cni}
\bibinfo{author}{H.~Mäntysaari}, \bibinfo{author}{B.~Schenke},
  \bibinfo{author}{C.~Shen}, \bibinfo{author}{P.~Tribedy},
  \bibinfo{journal}{Phys. Lett.} \bibinfo{volume}{B772} (\bibinfo{year}{2017})
  \bibinfo{pages}{681--686}. \href{http://arxiv.org/abs/1705.03177}{\tt
  arXiv:1705.03177}.
%Type = Article
\bibitem[{Jaiswal(2013)}]{Jaiswal:2013vta}
\bibinfo{author}{A.~Jaiswal}, \bibinfo{journal}{Phys. Rev.}
  \bibinfo{volume}{C88} (\bibinfo{year}{2013}) \bibinfo{pages}{021903}.
  \href{http://arxiv.org/abs/1305.3480}{\tt arXiv:1305.3480}.
%Type = Article
\bibitem[{Alqahtani et~al.(2018)Alqahtani, Nopoush, and
  Strickland}]{Alqahtani:2017mhy}
\bibinfo{author}{M.~Alqahtani}, \bibinfo{author}{M.~Nopoush},
  \bibinfo{author}{M.~Strickland}, \bibinfo{journal}{Prog. Part. Nucl. Phys.}
  \bibinfo{volume}{101} (\bibinfo{year}{2018}) \bibinfo{pages}{204--248}.
  \href{http://arxiv.org/abs/1712.03282}{\tt arXiv:1712.03282}.
%Type = Article
\bibitem[{Romatschke(2016)}]{Romatschke:2015gic}
\bibinfo{author}{P.~Romatschke}, \bibinfo{journal}{Eur. Phys. J.}
  \bibinfo{volume}{C76} (\bibinfo{year}{2016}) \bibinfo{pages}{352}.
  \href{http://arxiv.org/abs/1512.02641}{\tt arXiv:1512.02641}.
%Type = Article
\bibitem[{Grozdanov et~al.(2016)Grozdanov, Kaplis, and
  Starinets}]{Grozdanov:2016vgg}
\bibinfo{author}{S.~Grozdanov}, \bibinfo{author}{N.~Kaplis},
  \bibinfo{author}{A.~O. Starinets}, \bibinfo{journal}{JHEP}
  \bibinfo{volume}{07} (\bibinfo{year}{2016}) \bibinfo{pages}{151}.
  \href{http://arxiv.org/abs/1605.02173}{\tt arXiv:1605.02173}.
%Type = Article
\bibitem[{Kurkela et~al.(2018)Kurkela, Mazeliauskas, Paquet, Schlichting, and
  Teaney}]{Kurkela:2018wud}
\bibinfo{author}{A.~Kurkela}, \bibinfo{author}{A.~Mazeliauskas},
  \bibinfo{author}{J.-F. Paquet}, \bibinfo{author}{S.~Schlichting},
  \bibinfo{author}{D.~Teaney}  (\bibinfo{year}{2018}).
  \href{http://arxiv.org/abs/1805.01604}{\tt arXiv:1805.01604}.
%Type = Article
\bibitem[{Niemi and Denicol(2014)}]{Niemi:2014wta}
\bibinfo{author}{H.~Niemi}, \bibinfo{author}{G.~S. Denicol}
  (\bibinfo{year}{2014}). \href{http://arxiv.org/abs/1404.7327}{\tt
  arXiv:1404.7327}.
%Type = Article
\bibitem[{Alqahtani et~al.(2017)Alqahtani, Nopoush, and
  Strickland}]{Alqahtani:2016rth}
\bibinfo{author}{M.~Alqahtani}, \bibinfo{author}{M.~Nopoush},
  \bibinfo{author}{M.~Strickland}, \bibinfo{journal}{Phys. Rev.}
  \bibinfo{volume}{C95} (\bibinfo{year}{2017}) \bibinfo{pages}{034906}.
  \href{http://arxiv.org/abs/1605.02101}{\tt arXiv:1605.02101}.
%Type = Article
\bibitem[{Gallmeister et~al.(2018)Gallmeister, Niemi, Greiner, and
  Rischke}]{Gallmeister:2018mcn}
\bibinfo{author}{K.~Gallmeister}, \bibinfo{author}{H.~Niemi},
  \bibinfo{author}{C.~Greiner}, \bibinfo{author}{D.~H. Rischke},
  \bibinfo{journal}{Phys. Rev.} \bibinfo{volume}{C98} (\bibinfo{year}{2018})
  \bibinfo{pages}{024912}. \href{http://arxiv.org/abs/1804.09512}{\tt
  arXiv:1804.09512}.
%Type = Article
\bibitem[{Strickland et~al.(2018)Strickland, Noronha, and
  Denicol}]{Strickland:2017kux}
\bibinfo{author}{M.~Strickland}, \bibinfo{author}{J.~Noronha},
  \bibinfo{author}{G.~Denicol}, \bibinfo{journal}{Phys. Rev.}
  \bibinfo{volume}{D97} (\bibinfo{year}{2018}) \bibinfo{pages}{036020}.
  \href{http://arxiv.org/abs/1709.06644}{\tt arXiv:1709.06644}.
%Type = Article
\bibitem[{Pratt and Torrieri(2010)}]{Pratt:2010jt}
\bibinfo{author}{S.~Pratt}, \bibinfo{author}{G.~Torrieri},
  \bibinfo{journal}{Phys. Rev.} \bibinfo{volume}{C82} (\bibinfo{year}{2010})
  \bibinfo{pages}{044901}. \href{http://arxiv.org/abs/1003.0413}{\tt
  arXiv:1003.0413}.
%Type = Article
\bibitem[{Alqahtani et~al.(2017{\natexlab{a}})Alqahtani, Nopoush, Ryblewski,
  and Strickland}]{Alqahtani:2017jwl}
\bibinfo{author}{M.~Alqahtani}, \bibinfo{author}{M.~Nopoush},
  \bibinfo{author}{R.~Ryblewski}, \bibinfo{author}{M.~Strickland},
  \bibinfo{journal}{Phys. Rev. Lett.} \bibinfo{volume}{119}
  (\bibinfo{year}{2017}{\natexlab{a}}) \bibinfo{pages}{042301}.
  \href{http://arxiv.org/abs/1703.05808}{\tt arXiv:1703.05808}.
%Type = Article
\bibitem[{Alqahtani et~al.(2017{\natexlab{b}})Alqahtani, Nopoush, Ryblewski,
  and Strickland}]{Alqahtani:2017tnq}
\bibinfo{author}{M.~Alqahtani}, \bibinfo{author}{M.~Nopoush},
  \bibinfo{author}{R.~Ryblewski}, \bibinfo{author}{M.~Strickland},
  \bibinfo{journal}{Phys. Rev.} \bibinfo{volume}{C96}
  (\bibinfo{year}{2017}{\natexlab{b}}) \bibinfo{pages}{044910}.
  \href{http://arxiv.org/abs/1705.10191}{\tt arXiv:1705.10191}.
%Type = Article
\bibitem[{Almaalol et~al.(2018)Almaalol, Alqahtani, and
  Strickland}]{Almaalol:2018gjh}
\bibinfo{author}{D.~Almaalol}, \bibinfo{author}{M.~Alqahtani},
  \bibinfo{author}{M.~Strickland}  (\bibinfo{year}{2018}).
  \href{http://arxiv.org/abs/1807.04337}{\tt arXiv:1807.04337}.
%Type = Article
\bibitem[{Bozek et~al.(2016)Bozek, Broniowski, and Rybczy?ski}]{Bozek:2016kpf}
\bibinfo{author}{P.~Bozek}, \bibinfo{author}{W.~Broniowski},
  \bibinfo{author}{M.~Rybczy?ski}, \bibinfo{journal}{Phys. Rev.}
  \bibinfo{volume}{C94} (\bibinfo{year}{2016}) \bibinfo{pages}{014902}.
  \href{http://arxiv.org/abs/1604.07697}{\tt arXiv:1604.07697}.
%Type = Article
\bibitem[{Welsh et~al.(2016)Welsh, Singer, and Heinz}]{Welsh:2016siu}
\bibinfo{author}{K.~Welsh}, \bibinfo{author}{J.~Singer}, \bibinfo{author}{U.~W.
  Heinz}, \bibinfo{journal}{Phys. Rev.} \bibinfo{volume}{C94}
  (\bibinfo{year}{2016}) \bibinfo{pages}{024919}.
  \href{http://arxiv.org/abs/1605.09418}{\tt arXiv:1605.09418}.
%Type = Article
\bibitem[{Mäntysaari and Schenke(2017)}]{Mantysaari:2016how}
\bibinfo{author}{H.~Mäntysaari}, \bibinfo{author}{B.~Schenke},
  \bibinfo{journal}{Nucl. Part. Phys. Proc.} \bibinfo{volume}{289-290}
  (\bibinfo{year}{2017}) \bibinfo{pages}{457--460}.
  \href{http://arxiv.org/abs/1612.00041}{\tt arXiv:1612.00041}.
%Type = Article
\bibitem[{Bozek and Broniowski(2017)}]{Bozek:2017elk}
\bibinfo{author}{P.~Bozek}, \bibinfo{author}{W.~Broniowski},
  \bibinfo{journal}{Phys. Rev.} \bibinfo{volume}{C96} (\bibinfo{year}{2017})
  \bibinfo{pages}{014904}. \href{http://arxiv.org/abs/1701.09105}{\tt
  arXiv:1701.09105}.
%Type = Article
\bibitem[{Albacete et~al.(2018)Albacete, Petersen, and
  Soto-Ontoso}]{Albacete:2017ajt}
\bibinfo{author}{J.~L. Albacete}, \bibinfo{author}{H.~Petersen},
  \bibinfo{author}{A.~Soto-Ontoso}, \bibinfo{journal}{Phys. Lett.}
  \bibinfo{volume}{B778} (\bibinfo{year}{2018}) \bibinfo{pages}{128--136}.
  \href{http://arxiv.org/abs/1707.05592}{\tt arXiv:1707.05592}.
%Type = Article
\bibitem[{Romatschke(2018)}]{Romatschke:2018wgi}
\bibinfo{author}{P.~Romatschke}, \bibinfo{journal}{Eur. Phys. J.}
  \bibinfo{volume}{C78} (\bibinfo{year}{2018}) \bibinfo{pages}{636}.
  \href{http://arxiv.org/abs/1802.06804}{\tt arXiv:1802.06804}.
%Type = Article
\bibitem[{Kurkela et~al.(2018)Kurkela, Wiedemann, and Wu}]{Kurkela:2018qeb}
\bibinfo{author}{A.~Kurkela}, \bibinfo{author}{U.~A. Wiedemann},
  \bibinfo{author}{B.~Wu}  (\bibinfo{year}{2018}).
  \href{http://arxiv.org/abs/1805.04081}{\tt arXiv:1805.04081}.
%Type = Article
\bibitem[{Borghini et~al.(2018)Borghini, Feld, and Kersting}]{Borghini:2018xum}
\bibinfo{author}{N.~Borghini}, \bibinfo{author}{S.~Feld},
  \bibinfo{author}{N.~Kersting}, \bibinfo{journal}{Eur. Phys. J.}
  \bibinfo{volume}{C78} (\bibinfo{year}{2018}) \bibinfo{pages}{832}.
  \DOIprefix\doi{10.1140/epjc/s10052-018-6313-z}.
  \href{http://arxiv.org/abs/1804.05729}{\tt arXiv:1804.05729}.
%Type = Article
\bibitem[{Xu and Greiner(2005)}]{Xu:2004mz}
\bibinfo{author}{Z.~Xu}, \bibinfo{author}{C.~Greiner}, \bibinfo{journal}{Phys.
  Rev.} \bibinfo{volume}{C71} (\bibinfo{year}{2005}) \bibinfo{pages}{064901}.
  \href{http://arxiv.org/abs/hep-ph/0406278}{\tt arXiv:hep-ph/0406278}.
%Type = Article
\bibitem[{Lin et~al.(2005)Lin, Ko, Li, Zhang, and Pal}]{Lin:2004en}
\bibinfo{author}{Z.-W. Lin}, \bibinfo{author}{C.~M. Ko}, \bibinfo{author}{B.-A.
  Li}, \bibinfo{author}{B.~Zhang}, \bibinfo{author}{S.~Pal},
  \bibinfo{journal}{Phys. Rev.} \bibinfo{volume}{C72} (\bibinfo{year}{2005})
  \bibinfo{pages}{064901}. \href{http://arxiv.org/abs/nucl-th/0411110}{\tt
  arXiv:nucl-th/0411110}.
%Type = Article
\bibitem[{Ghiglieri et~al.(2018)}]{nloetas}
\bibinfo{author}{J.~Ghiglieri}, et~al., \bibinfo{journal}{Quark Matter 2018,
  these proceedings}  (\bibinfo{year}{2018}).
%Type = Article
\bibitem[{Mace et~al.(2018)}]{mace}
\bibinfo{author}{M.~Mace}, et~al., \bibinfo{journal}{Quark Matter 2018, these
  proceedings}  (\bibinfo{year}{2018}).
%Type = Article
\bibitem[{Greif et~al.(2017)Greif, Greiner, Schenke, Schlichting, and
  Xu}]{Greif:2017bnr}
\bibinfo{author}{M.~Greif}, \bibinfo{author}{C.~Greiner},
  \bibinfo{author}{B.~Schenke}, \bibinfo{author}{S.~Schlichting},
  \bibinfo{author}{Z.~Xu}, \bibinfo{journal}{Phys. Rev.} \bibinfo{volume}{D96}
  (\bibinfo{year}{2017}) \bibinfo{pages}{091504}.
  \href{http://arxiv.org/abs/1708.02076}{\tt arXiv:1708.02076}.
%Type = Article
\bibitem[{{Y-J. Lee (ALEPH Collaboration)}(2018)}]{aleph}
\bibinfo{author}{{Y-J. Lee (ALEPH Collaboration)}}, \bibinfo{journal}{Quark
  Matter 2018, these proceedings}  (\bibinfo{year}{2018}).
%Type = Article
\bibitem[{{J. Onderwaater, (ZEUS Collaboration)}(2018)}]{zeus}
\bibinfo{author}{{J. Onderwaater, (ZEUS Collaboration)}},
  \bibinfo{journal}{Quark Matter 2018, these proceedings}
  (\bibinfo{year}{2018}).
%Type = Article
\bibitem[{Shen et~al.(2016)Shen, Park, Paquet, Denicol, Jeon, and
  Gale}]{Shen:2016egw}
\bibinfo{author}{C.~Shen}, \bibinfo{author}{C.~Park}, \bibinfo{author}{J.-F.
  Paquet}, \bibinfo{author}{G.~S. Denicol}, \bibinfo{author}{S.~Jeon},
  \bibinfo{author}{C.~Gale}, \bibinfo{journal}{Nucl. Phys.}
  \bibinfo{volume}{A956} (\bibinfo{year}{2016}) \bibinfo{pages}{741--744}.
  \href{http://arxiv.org/abs/1601.03070}{\tt arXiv:1601.03070}.
%Type = Article
\bibitem[{Acharya et~al.(2018)}]{Acharya:2017okq}
\bibinfo{author}{S.~Acharya}, et~al. (\bibinfo{collaboration}{ALICE}),
  \bibinfo{journal}{Phys. Lett.} \bibinfo{volume}{B783} (\bibinfo{year}{2018})
  \bibinfo{pages}{95--113}. \href{http://arxiv.org/abs/1712.05603}{\tt
  arXiv:1712.05603}.
%Type = Article
\bibitem[{Khachatryan et~al.(2016)}]{Khachatryan:2016xxp}
\bibinfo{author}{V.~Khachatryan}, et~al. (\bibinfo{collaboration}{CMS}),
  \bibinfo{journal}{Phys. Lett. B}  (\bibinfo{year}{2016}).
  \href{http://arxiv.org/abs/1611.01510}{\tt arXiv:1611.01510}.
%Type = Misc
\bibitem[{{CMS Physics Analysis Summary, CMS PAS HIN-16-023}(2017)}]{CMS5TeV}
\bibinfo{author}{{CMS Physics Analysis Summary, CMS PAS HIN-16-023}},
  \bibinfo{howpublished}{\url{http://cds.cern.ch/record/2244680/files/HIN-16-023-pas.pdf}},
  \bibinfo{year}{2017}.
%Type = Misc
\bibitem[{{Indranil Das and Antoine Lardeux (ALICE
  Collaboration)}(2017)}]{ALICE5TeV}
\bibinfo{author}{{Indranil Das and Antoine Lardeux (ALICE Collaboration)}},
  \bibinfo{howpublished}{Quark Matter 2017, \url{https://goo.gl/aAktfl}},
  \bibinfo{year}{2017}.
%Type = Article
\bibitem[{Aaboud et~al.(2018)}]{Aaboud:2017cif}
\bibinfo{author}{M.~Aaboud}, et~al. (\bibinfo{collaboration}{ATLAS}),
  \bibinfo{journal}{Eur. Phys. J.} \bibinfo{volume}{C78} (\bibinfo{year}{2018})
  \bibinfo{pages}{171}. \href{http://arxiv.org/abs/1709.03089}{\tt
  arXiv:1709.03089}.
%Type = Article
\bibitem[{Emerick et~al.(2012)Emerick, Zhao, and Rapp}]{Emerick:2011xu}
\bibinfo{author}{A.~Emerick}, \bibinfo{author}{X.~Zhao},
  \bibinfo{author}{R.~Rapp}, \bibinfo{journal}{Eur. Phys. J.}
  \bibinfo{volume}{A48} (\bibinfo{year}{2012}) \bibinfo{pages}{72}.
  \href{http://arxiv.org/abs/1111.6537}{\tt arXiv:1111.6537}.
%Type = Article
\bibitem[{Strickland(2011)}]{Strickland:2011mw}
\bibinfo{author}{M.~Strickland}, \bibinfo{journal}{Phys. Rev. Lett.}
  \bibinfo{volume}{107} (\bibinfo{year}{2011}) \bibinfo{pages}{132301}.
  \href{http://arxiv.org/abs/1106.2571}{\tt arXiv:1106.2571}.
%Type = Article
\bibitem[{Strickland and Bazow(2012)}]{Strickland:2011aa}
\bibinfo{author}{M.~Strickland}, \bibinfo{author}{D.~Bazow},
  \bibinfo{journal}{Nucl. Phys.} \bibinfo{volume}{A879} (\bibinfo{year}{2012})
  \bibinfo{pages}{25--58}. \href{http://arxiv.org/abs/1112.2761}{\tt
  arXiv:1112.2761}.
%Type = Article
\bibitem[{Krouppa et~al.(2015)Krouppa, Ryblewski, and
  Strickland}]{Krouppa:2015yoa}
\bibinfo{author}{B.~Krouppa}, \bibinfo{author}{R.~Ryblewski},
  \bibinfo{author}{M.~Strickland}, \bibinfo{journal}{Phys. Rev.}
  \bibinfo{volume}{C92} (\bibinfo{year}{2015}) \bibinfo{pages}{061901}.
  \href{http://arxiv.org/abs/1507.03951}{\tt arXiv:1507.03951}.
%Type = Article
\bibitem[{Krouppa and Strickland(2016)}]{Krouppa:2016jcl}
\bibinfo{author}{B.~Krouppa}, \bibinfo{author}{M.~Strickland},
  \bibinfo{journal}{Universe} \bibinfo{volume}{2} (\bibinfo{year}{2016})
  \bibinfo{pages}{16}. \href{http://arxiv.org/abs/1605.03561}{\tt
  arXiv:1605.03561}.
%Type = Article
\bibitem[{Du et~al.(2017)Du, Rapp, and He}]{Du:2017qkv}
\bibinfo{author}{X.~Du}, \bibinfo{author}{R.~Rapp}, \bibinfo{author}{M.~He},
  \bibinfo{journal}{Phys. Rev.} \bibinfo{volume}{C96} (\bibinfo{year}{2017})
  \bibinfo{pages}{054901}. \href{http://arxiv.org/abs/1706.08670}{\tt
  arXiv:1706.08670}.
%Type = Article
\bibitem[{Krouppa et~al.(2017)Krouppa, Ryblewski, and
  Strickland}]{Krouppa:2017lsw}
\bibinfo{author}{B.~Krouppa}, \bibinfo{author}{R.~Ryblewski},
  \bibinfo{author}{M.~Strickland}, \bibinfo{journal}{Nucl. Phys.}
  \bibinfo{volume}{A967} (\bibinfo{year}{2017}) \bibinfo{pages}{604--607}.
  \href{http://arxiv.org/abs/1704.02361}{\tt arXiv:1704.02361}.
%Type = Article
\bibitem[{Krouppa et~al.(2018)Krouppa, Rothkopf, and
  Strickland}]{Krouppa:2017jlg}
\bibinfo{author}{B.~Krouppa}, \bibinfo{author}{A.~Rothkopf},
  \bibinfo{author}{M.~Strickland}, \bibinfo{journal}{Phys. Rev.}
  \bibinfo{volume}{D97} (\bibinfo{year}{2018}) \bibinfo{pages}{016017}.
  \href{http://arxiv.org/abs/1710.02319}{\tt arXiv:1710.02319}.
%Type = Article
\bibitem[{Ferreiro and Lansberg(2018)}]{Ferreiro:2018wbd}
\bibinfo{author}{E.~G. Ferreiro}, \bibinfo{author}{J.-P. Lansberg},
  \bibinfo{journal}{JHEP} \bibinfo{volume}{10} (\bibinfo{year}{2018})
  \bibinfo{pages}{094}. \DOIprefix\doi{10.1007/JHEP10(2018)094}.
  \href{http://arxiv.org/abs/1804.04474}{\tt arXiv:1804.04474}.

\end{thebibliography}






% \end{thebibliography}

\end{document}